# Data Source Selection for Information Integration in Big Data Era

Yiming Lin, Hongzhi Wang, Jianzhong Li, Hong Gao
Harbin Institute of Technology, Harbin, China
YimingLin_0426@hotmail.com {wangzh,lijzh,honggao}@hit.edu.cn

*Abstract*—In Big data era, information integration often requires abundant data extracted from massive data sources. Due to a large number of data sources, data source selection plays a crucial role in information integration, since it is costly and even impossible to access all data sources. Data Source selection should consider both efficiency and effectiveness issues. For efficiency, the approach should achieve high performance and be scalability to fit large data source amount. From effectiveness aspect, data quality and overlapping of sources are to be considered, since data quality varies much from data sources, with significant differences in the accuracy and coverage of the data provided, and the overlapping of sources can even lower the quality of data integrated from selected data sources. In this paper, we study source selection problem in *Big Data Era* and propose methods which can scale to datasets with up to millions of data sources and guarantee the quality of results.

Motivated by this, we propose a new object function taking the expected number of true values a source can provide as a criteria to evaluate the contribution of a data source. Based on our proposed index we present a scalable algorithm and two pruning strategies to improve the efficiency without sacrificing precision. Experimental results on both real world and synthetic data sets show that our methods can select sources providing a large proportion of true values efficiently and can scale to massive data sources.

## I. Introduction

Information integration is to fuse information from heterogeneous sources with differing conceptual, contextual and typographical representations. Information integration in big data era is critical, since variety is an essential feature of big data and the data-centric applications often require data from heterogeneous data sources. As an example, the construction of knowledge bases from big data on the web, such as DBpedia, Freebase, requires the integration of data from massive autonomous, heterogeneous data sources distributed on the web.

In big data era, data sources have following new features.

- *Volume*: The big data era has witnessed that not only can a data source contain a large volume of data, but also the number of data sources grows at an unprecedented scale. This data source number may scale to tens of million[1], which is much larger than that of data sources in traditional information integration.
- *Variety*: Data quality varies much from source to source even in the same domain, with significant differences in the accuracy and coverage of the data provided.
- *Overlap*: The data in a large amount of data sources have a great chance to contain overlapping information due to the autonomy of data sources. It makes overlapping more serious that some data sources may copy information from others.

Due to volume, data source selection plays a crucial role in information integration, since it is costly and even impossible to access all the data sources. These new features bring the challenges in efficiency and effectiveness aspects. First, different from traditional information integration, source selection in big data era has to handle massive data sources. Thus, data source selection algorithms should be efficient and scalable. However, existing techniques such as [1], [2] own high computational complexity and can hardly scale to millions of data sources. The scalabilty poses a big challenge. Second, the large amount of autonomous data sources cause imbalance data quality and complex overlapping relationship among them. They have to be handled during data source selection due to the effectiveness. To obtain high-quality results from a small share of selected data sources, the quality and coverage of selected data sources should be considered. As a result, the challenge is to ensure the quality and coverage of the selected data sources.

We use an example to illustrate such challenges.

*Example 1:* In the motivating example in Table I, we ask the question "where is the location of the headquarters of the following companies, AT&T, Google, Microsoft, IBM and Apple?" ($q_1$). There are five data items, the location of AT&T, Google, Microsoft, IBM and Apple. Data from seven sources, $S_1$ to $S_7$, provides candidates for the query.

Overlapping among sources is significant: for $AT\&T$, $S_1$, $S_2$, $S_3$, $S_6$ provide the same value DA. And the location information from different data sources for the same company may conflict and some sources may contain missing values. The location of AT&T is described by three values {DA, TE, NY} and there is a missing value on $S_5$.

For data source selection for $q_1$, without the consideration of data quality, all the data sources have to be accessed. On one hand, the accessing of unnecessary data sources that are query-irrelevant or provide little answers to the query lowers efficiency. On the other hand, low-quality data sources, such as conflicting in data sources, results in low-quality query results.

Each attribute value in Table I comes with a probability that the value is a truth for the data item it describes. The probability value can be derived by many truth discovery techniques [3], [4] and here we adopt [3]. As for IBM, NY owns the highest probability, .58 while WA is .34, both of which differ not very much. And we can not deem the value with the highest true probability to be the truth for sure.

TABLE I. MOTIVATING EXAMPLE

|       | AT&T    | Google  | Microsoft | IBM     | Apple   |
|-------|---------|---------|-----------|---------|---------|
| $S_1$ | DA(.64) | CA(.74) | WA(.99)   | WA(.34) | CA(.93) |
| $S_2$ | DA(.64) | LA(.26) | WA(.99)   | NY(.58) | CA(.93) |
| $S_3$ | DA(.64) |         | TX(.00)   | NY(.58) | CA(.93) |
| $S_4$ | TE(.23) | LA(.26) | WA(.99)   | BS(.08) | CA(.93) |
| $S_5$ |         | CA(.74) | BJ(.01)   | NY(.58) | NY(.02) |
| $S_6$ | DA(.64) | CA(.74) | BJ(.01)   | NY(.58) |         |
| $S_7$ | NY(.13) | CA(.74) | WA(.99)   | NY(.58) | WA(.05) |

Due to the importance, data source selection draws attention in literature. [1] selected a set of sources to maximize the profit taking the monotonicity property into consideration and proposed a heuristic random approach. Their focus is effectiveness, not scalability. Thus, when executing algorithms on datasets with up to millions of data sources, it is extremely time-consuming. [2] proposed a data integration technique with dependent sources under the assumption that all values provided by sources are true. Even though the proposed algorithm can estimate the coverage and select sources effectively, it does not consider data quality and owns over-cubic time complexity, which prevents the algorithm from scaling to massive data sources.

Motivated by these, we propose efficient source selection approaches which can scale up to millions of sources. At the same time, the quality of results is guaranteed. We solve this problem in two aspects of quality and efficiency.

Considering data quality, we tend to choose a source that provides more "true" values, since for the same data item, different sources can provide conflicting values, among which not all are true. Our goal is to select a subset of data sources which provide the most truths describing the given data items. We call the process of resolving conflicting values by truth discovery, which has been widely studied [3], [4]. We incorporate *truth discovery* techniques into source selection to improve the data quality of selected data sources. To achieve this goal, we devise the probability of a value being true and further define a *probabilistic coverage model* to derive the quality metric of a data source. That is, we take the *expected number* of true values returned by a source as a criteria. Considering this criteria, we formulate the data source problem as two optimization problems from different aspects, both of which are proven to be NP-hard.

Regarding efficiency and scalability, the computational steps of quality metric considering source overlapping is exponential w.r.t the number of data sources. We first present a quadratic algorithm to devise quality metric with the support of a sophisticated index and then describe a polynomial-time approximate algorithm for source selection problem. We also exploit the opportunities for efficiency improvement and propose two pruning strategies in both source-level and value-level to accelerate the source selection algorithm furthermore. Experimental results on both real-world and synthetic datasets show that our techniques can scale to millions of data sources efficiently.

In this paper, we make the following contributions.

First, we propose a probabilistic coverage model for data source selection from massive autonomous data sources. This model takes data quality and source overlapping into consideration. (Section II)

Second, under the proposed model, we formalize the data source selection problem into two optimization problems and prove their hardness. To solve the problem efficiently, we design an index and devise an efficient algorithm with performance guarantee, which enables the approach to scale to massive sources. (Section III)

Third, to further improve the efficiency of the algorithm, we propose two efficient pruning strategies. Both are with theoretical guarantee and will not sacrifice the quality of results much. (Section IV)

Finally, we conduct a comprehensive evaluation of our technique against real and synthetic data sets. We empirically show that even with data quality problems, our methods can select sources that provide a large proportion of true values efficiently and can scale to millions of data sources. (Section V)

The remainder of our paper is structured as follows.

Section II introduces basic notions and proposes a new probabilistic model for data source selection, then defines the problems formally. Section III describes a basic solution and a scalable index-based algorithm. Section IV presents two pruning strategies. Section V reports experimental results with analysis. Section VI discusses related work, and Section VII concludes the whole paper.

## II. PROBLEM DEFINITION

In this section, we first define basic notions used in our paper formally as well as the quality metric of a data source; next we formulate the problem of source selection.

### A. Basic Notions and Quality Metric

*Definition 1:* (Data source) Consider a set of data sources $\Omega$. Each source $S \in \Omega$ is associated with a cost $c_S$ and contains a set of tuples $\mathcal{T}$, $S = \{t_1, t_2, ..., t_{|\mathcal{T}|}\}$. Each tuple $t \in \mathcal{T}$ consists of a set of attributes value $\mathcal{ATT} = \{v_1, v_2, ..., v_{|\mathcal{ATT}|}\}$.

Please note that the cost of integrating a data source comes from two aspects. On one hand, some data sources, such as WDT for weather data, charges for its data[1]. On the other hand, even if some sources are free, data preparations for integration, such as resolving conflicting values, cleaning the data and mapping heterogeneous data items require resources. In our paper, given a source set $\mathcal{S}$, the total cost $c_\mathcal{S} = \sum_{i=1}^{|\mathcal{S}|} c_{S_i}$.

*Definition 2:* (Data item) Consider a data item domain $\mathcal{D}$. A data item $D \in \mathcal{D}$ represents a particular aspect of a real-world entity. For each $D \in \mathcal{D}$, a source $S$ can (but not necessarily) provide an attribute value.

In motivating example, a data item is "the location of the headquaters of Google". Among all distinct values describing a data item, we assume that only one is true, and the rest are false. Note that in some cases one data item has multiple true values, e.g., one musician may have multiple albums. In this paper, we omit this case and the assumption of one-true-value already fits many real-world applications.

As surveyed in [5], data in many sources are of low and unstable quality. Even for some domains that most people consider as highly reliable, it is observed that a large amount of inconsistency exists. Thus, when selecting data sources for information integration, in many cases, it is infeasible to determine whether a value returned from a source is true or false for sure. Motivated by this, we leverage the uncertainty introduced in [6] and assume that each value is associated with a probability of correctness.

*Definition 3:* (Uncertainty) Given $\Omega$ and $\mathcal{D}$, each attribute value $v$ associates a truth probability, denoted by $P(v)$, which is the probability of a value $v$ being true.

For the computational methods of truth probability, we handle it as black box. And we adopt the techniques proposed by [3] to compute $P(v)$ of $v$ for the following reasons. First, it considers both source accuracy and copying, and thus can compute a truth probability with a high precision. Second, it could be executed off-line efficiently.

*Definition 4:* (Query) Let $Q$ be a query and $S$ be a data source, we define by $Q(S)$ the answer set returned by $S$.

Even though a query may have various forms, basically, from the aspect of data source accessing, a query could be converted into a series of selection operations, while other operations are executed in the mediator. Thus, a query could be converted a series of queries for item selection from data sources.

Continuing with motivating example, for the query $q_1$, which queries for five data items, the location of AT&T, Google, Microsoft, IBM and Apple. For source $S_2$, $Q(S_2) = \{DA, LA, WA, NY, CA\}$.

Next, we consider quality metric. As widely used in [7], [8], *coverage* of a source $S$ reflects the proportion of answers returned by $S$ in the whole set $\Omega$. Taking uncertainty into account, each answer comes with a truth probability and we introduce the *expectation of answers* in the quality metric. Then we define our probabilistic coverage model as follows.

*Definition 5:* (Probabilistic Coverage Model) Let $S$ be a source, $\mathbb{S} \subseteq \Omega$ be a subset of sources and $Q$ be a query. The probabilistic of coverage for $S$ and $\mathbb{S}$ are defined as follows,

$$Cov(S) = \sum_{v \in Q(S)} P(v) \qquad (1)$$
$$Cov(\mathbb{S}) = \sum_{v \in (\bigcup_{S \in \mathbb{S}} Q(S))} P(v) \qquad (2)$$

$Cov(S)$ is the expected number of answers returned by $S$ and it should be devised as $\sum_{v \in Q(S)} P(v) \times 1$, where "1" denotes that each value contributes one number to the final answer set. Then we derive the expectation by multiplying "1" with its true probability $P(v)$. $\bigcup_{S \in \mathbb{S}} Q(S)$ denotes the answer set returned by $\mathbb{S}$. Please note that each value in this answer set is *distinct*. In the remainder of our paper, we will use *contribution* of source $S$ to refer to the probabilistic coverage of $S$, so does $\mathbb{S}$.

### B. Problems

In this subsection we formulate our problem.

Given $\mathcal{D}$, for a data source $S$, it comes with a cost $c_S$ and owns a contribution $Cov(S)$. The combination of these two factors has two possibilities. One is to pursue the maximal contribution of selected data sources with the cost under the fixed upper bound. The other requires the maximal contribution of selected data sources, i.e. equaling to that of all data sources, with minimizing the cost as the goal. These two possibilities have their suitable scenarios, respectively. We define them as MaxContribution problem and MinCost problem as follows.

MaxContribution In some scenarios, information integration system could tolerate incomplete results but the cost could not exceed the budget. For such scenarios, we define MaxContribution problem with the upper bound of cost as the constraint for total cost and the contribution of the selected sources as the optimization goal. Formally, the problem is defined as follows.

*Problem 1:* Given a query $Q$ and a data source set $\Omega = \{S_1, S_2, ..., S_n\}$ and a cost upper bound $L$, find a set of data sources $\mathbb{S} \in \Omega$, such that

$$\max Cov(\mathbb{S})$$
$$\text{Subject to: } \sum_{S \in \mathbb{S}} c_S \leq L \qquad (3)$$

MinCost In some scenarios, *all answers* of a given query are required. Since the source sets satisfying this condition may have multiple choices, the optimization choice is to minimize the total cost of selected data sources. The problem is defined as follows.

*Problem 2:* Given a query $Q$ and a data source set $\Omega = \{S_1, S_2, ..., S_n\}$, find a set of data sources $\mathbb{S} \in \Omega$, such that

$$\min \sum_{S \in \mathbb{S}} c_S$$
$$\text{Subject to: } Cov(\mathbb{S}) = Cov(\Omega) \qquad (4)$$

*Theorem 1:* Both **MaxContribution** and **MinCost** problems are NP-hard problems.

*Proof 1:* We prove the NP-hardness for MaxContribution problem by reducing knapsack problem [9] to it.

For an instance $I_{max}$ of knapsack problem with the capability as $L$ and an object set $O=\{o_1, o_2, \cdots, o_m\}$ with the price and weight the object $o_i \in O$ denoted by $p_i$ and $w_i$, respectively. We construct an instance $I'_{max}$ of MaxContribution problem with cost upper bound $L$ and a source collection $\Omega=\{S_1, S_2, \cdots, S_n\}$ and a domain of values $V=\{v_1, v_2, \cdots, v_n\}$. For $S_i \in \Omega$, $S_i=\{v_i\}$ and its cost is $w_i$. For each $v_i \in V$, its truth probability is $p_i$.

Then we construct a solution of $I_{max}$, denoted by $R$, from $\mathbb{S}$ as the solution of $I'_{max}$ by selecting the $i$th object for each $S_i \in \mathbb{S}$.

We attempt to prove that $R$ is the optimal solution of $I_{max}$. Given $\mathbb{S}$ as the optimal solution for $I'_{max}$, it satisfies that $\sum_{S_i \in \mathbb{S}} w_i \leq L$ and $Cov(\mathbb{S})=\sum_{S_i \in \mathbb{S}} Cov(S_i)=\sum_{S_i \in \mathbb{S}} p_i$ is maximal. With the construction approach of the solution, $\sum_{S_i \in \mathbb{S}} w_i \leq L$ and a maximal $Cov(\mathbb{S})=\sum_{S_i \in \mathbb{S}} p_i$ imply $\sum_{o_i \in R} w_i \leq L$ and a maximal $\sum_{o_i \in R} p_i$, respectively, which are exactly the constraint and optimal goal of Knapsack.

Clearly, the reduction could be accomplished in polynomial time. Since Knapsack is an NP-hard problem [9], MaxContribution problem is an NP-hard problem.

We prove the NP-hardness of MinCost problem by reducing set cover problem [9] to it.

Given an instance $I_{min}$ of the set cover problem with $U$ as the universal set and $\mathfrak{S}=\{S_1, S_2, \cdots, S_n\}$, we construct an instance $I'_{min}$ for MinCost problem with $\Omega=\mathfrak{S}$ and the value set $V=U$. For each $v_i \in V$, the truth probability $p_i=2^i$, and for each $S_i$, $c_i=1$.

Then we construct a solution of $I_{min}$, denoted by $R$, from $\mathbb{S}$ as the solution of $I'_{min}$ by selecting the $S_i$ for each $S_i \in \mathbb{S}$.

We attempt to prove that $R$ is the optimal solution of $I_{min}$. Given $\mathbb{S}$ as the optimal solution for $I'_{min}$, it satisfies that $Cov(\mathbb{S})=Cov(\Omega)$ and $\sum_{S_i \in \mathbb{S}} c_i$ is minimal. Since $p_i=2^i$, $Cov(\mathbb{S})=Cov(\Omega)=\sum_{i=1}^n 2^i$. Therefore, $\mathbb{S}=\bigcup_{S_i \in \mathbb{S}} S_i=\Omega=U$, which is the constraint of set cover problem. Since the cost of each $S_i$ is 1, $\sum_{S_i \in \mathbb{S}} c_{S_i}=|\mathbb{S}|$. Thus the minimal $\sum_{S_i \in \mathbb{S}} c_{S_i}$ implies a minimal $|\mathbb{S}|$, which is the optimization goal of set cover problem.

Since the reduction could be accomplished in polynomial time and set cover is an NP-hard problem [9], MinCost problem is an NP-hard problem.

## III. SOURCE SELECTION

In this section, we start with an overview of solutions of source selection in Section III-A; then we present method to compute coverage model efficiently defined in Definition 5 (Section III-B; finally we show our source algorithms (Section III-C).

### A. Overview

Recall that we formulate source selection problem into two problems, MaxContribution and MinCost. Intuitively, both seek to select a source set $\mathbb{S}$ to maximize $Cov(\mathbb{S})$ and minimize $c_\mathbb{S}$. Thus, there remains two problems to solve:

1) Given source set $\mathbb{S}$, how to compute $Cov(\mathbb{S})$ efficiently?

2) How to select sources efficiently to meet the constraints of two problems?

Correspondingly, our proposed approaches to source selection is 2-phase:

<u>Computation of Coverage Model:</u> Given a source set $\mathbb{S}$, $Cov(\mathbb{S})$ denotes the expected number of answers returned by $\mathbb{S}$. Due to source overlapping, that is, different sources could provide the same answers for given query, we show that computing for $Cov(\mathbb{S})$ requires exponential steps in Section III-B. And we propose a polynomial method to compute it precisely leveraging a sophiscated index structure.

<u>Source Selection:</u> In Section II-B, both problems are proven to be NP-hard. Since no efficient exact solutions to NP-hard problems are known, we develop greedy-based algorithms, which have approximation bound.

The greedy selection is that in each iteration, we tend to select a source with more new contribution and less cost. Given a selected data source set $\mathbb{S}$, we select such a source $S$ that:

$$S = argmax_{S \in \Omega \setminus \mathbb{S}}(Cov(S \cup \mathbb{S}) - Cov(\mathbb{S}))/c_S \quad (5)$$

In the remainder of paper, we use the term *selection ratio* to refer to $(Cov(S \cup \mathbb{S}) - Cov(\mathbb{S}))/c_S$. MaxContribution and MinCost are both greedy-based and share the same framework. They only differ in the stop condition: for MaxContribution, the stop condition is that the cost of selection sources reaches the upper bound of cost to meet the cost constraint; while MinCost algorithm stops when no data source could provide new results to meet the constraint of $Cov(\mathbb{S}) = Cov(\Omega)$.

### B. Computing Coverage Model

In this subsection, our goal is to compute coverage model efficiently. We first present a basic solution which requires exponential steps, and then we propose a polynomial method based on a well-designed index structure.

Recall that in *Overview* part, the final goal in each iteration is to select a source with maximum selection ratio. Thus in one iteration, given selected source set $\mathbb{S}$, for each $S \notin \mathbb{S}$, we compute $(Cov(S \cup \mathbb{S}) - Cov(\mathbb{S}))/c_S$. Since $c_S$ is known, we only need to calculate $Cov(S \cup \mathbb{S}) - Cov(\mathbb{S})$ for each source. We denote $Cov(S \cup \mathbb{S}) - Cov(\mathbb{S})$ as $\widehat{Cov}(S)$, thus the goal in each iteration is to select such a source $S$ that

$$S = argmax_{S \in \Omega \setminus \mathbb{S}} \widehat{Cov}(S)/c_S \quad (6)$$

Here we use the term *irreplaceable contribution* to refer to $\widehat{Cov}(S)$, which denotes the expected number of answers returned by $S$ but not by any source in $\mathbb{S}$.

Obviously, the computational complexity for $Cov(.)$ and $\widehat{Cov}(S)$ is the same. Instead of computing $Cov(.)$, we turn to discuss the methods to devise $\widehat{Cov}(S)$ efficiently.

<u>Basic solution:</u> We introduce the methods proposed by [2] as the basic solution, which applies inclusion-exclusion principle to obtain the irreplaceable contribution of $S$.

$$\widehat{Cov}(S) = Cov(S) - \sum_{i=1}^{|\mathbb{S}|} (-1)^{i-1} \sum_{\mathbb{S}_o \subseteq \mathbb{S}, |\mathbb{S}_o|=i} Cov(S \cap \mathbb{S}_o)$$

Here $\mathbb{S}_o$ is a subset of $\mathbb{S}$ and we enumerate its cardinality from 1 to $|\mathbb{S}|$. Next we use an example to illustrate the basic solution.

*Example 2:* Continuing with the motivating example in Table I, we assume that source $S_1$ and $S_2$ have been selected Then we consider probing $S_6$. By Equation 7 we calculate the irreplaceable contribution of $S_6$ as $\widehat{Cov}(S_6) = Cov(S_6) - Cov(S_1 \cap S_6) - Cov(S_2 \cap S_6) + Cov(S_1 \cap S_2 \cap S_6) = (.64+.74+.01+.58)-(.64+.74)-(.64+.58)+(.64) = .01$.

Note that computing $\widehat{Cov}(S)$ is non-trivial, which requires exponential computational steps. How to obtain it efficiently is critical to the performance of source selection.

<u>Polynomial solution leveraging index:</u> $\widehat{Cov}(S)$ denotes the expected number of true answers provided by $S$ but not covered by any selected source. Instead of computational method in Equation 7, we can compute it by accumulating the truth probability of each value returned *only* by $S$. Given a selected source set $\mathbb{S}$, we have:

$$\widehat{Cov}(S) = \sum_{v \in (Q(S)-Q(S \cap \mathbb{S}))} P(v) \quad (7)$$

TABLE II. INDEX STRUCTURE

| entry | corresponding sources | probability |
|---|---|---|
| AT&T.DA | $S_1, S_2, S_3, S_6$ | .64 |
| AT&T.TE | $S_4$ | .23 |
| AT&T.NY | $S_7$ | .13 |
| Google.CA | $S_1, S_5, S_6, S_7$ | .74 |
| Google.LA | $S_2, S_4$ | .26 |
| MS.WA | $S_1, S_2, S_4, S_7$ | .99 |
| MS.BJ | $S_5, S_6$ | .01 |
| MS.TX | $S_3$ | .00 |
| IBM.NY | $S_2, S_3, S_5, S_6, S_7$ | .58 |
| IBM.WA | $S_1$ | .34 |
| IBM.BS | $S_4$ | .08 |
| Apple.CA | $S_1, S_2, S_3, S_4$ | .93 |
| Apple.WA | $S_7$ | .05 |
| Apple.NY | $S_5$ | .02 |

where $Q(S) - Q(S \cap \mathbb{S})$ is the set of distinct answers returned by $S$ not by $\mathbb{S}$. We accumulate truth probability of such value to obtain irreplaceable contribution of $S$.

*Example 3:* Continuing with the motivating example in Table I, we assume that $S_1$ and $S_2$ have been selected, that is, $\mathbb{S} = \{S_1, S_2\}$. Next we consider $S_4$ containing five values AT&T.TE, Google.LA, MS.WA, IBM.BS and Apple.CA. Among them, Google.LA, MS.WA and Apple.CA are also covered by $\mathbb{S}$. Thus, we only need to accumulate truth probability of AT&T.TE(.23) and IBM.BS(.08). Accordingly the irreplaceable contribution of $S_4$ is calculated as $\widehat{Cov}(S_4) = .23 + .08 = .31$.

In order to devise $\widehat{Cov}(S)$, we need to access those values returned by $S$ but not by selected source set $\mathbb{S}$ quickly. Thus, we facilitate above process by designing an index structure.

*Definition 6:* Given a set of data items $\mathcal{D}$ and a set of data sources $\mathcal{S}$, the index for $\mathcal{D}$ and $\mathcal{S}$ contains a set of entries $\mathcal{E}$. For each entry $E \in \mathcal{E}$:

1) E corresponds to a value $D_E.v_E$, where $D_E \in \mathcal{D}$ and $v_E$ is a value provided for $D_E$;

2) E is associated with a truth probability $P(v_E)$;

3) the entry E contains data sources that provide value $v_E$ for $D_E$.

Next we use an example to illustrate the index.

*Example 4:* Table II shows the index built for data in Table I. Consider data item $AT\&T$, there are three values DA,TE and NY provided for it. Accordingly, we build three entries $AT\&T.DA$, $AT\&T.TE$ and $AT\&T.NY$. As for entry $AT\&T.DA$, it contains sources that provide DA for AT&T, $\{S_1, S_2, S_3, S_6\}$. And it is associated with a truth probability .64 denoting that $AT\&T.DA$ owns .64 probability being a truth.

According to Equation 7, in each round, when a source $S$ is selected, those values covered by $S$ should never be accessed in the following rounds by other sources to avoid redundancy. Accordingly, in order to facilitate the process above, when we implement algorithms on inverted lists, we build a flag bit for each entry in main memory to record whether the value corresponding to the entry is selected. In one round, if a source $S$ is selected, for each value provided by $S$, we set the flag bit of the entry corresponding to such values to be true, which means that such entry will never be accessed in later rounds since the values corresponding to the entries are selected.

Next we show the algorithm *Selectmax* to explicit the process of selecting a source with the maximum selection ratio in Algorithm 1.

First we set an array *count*, where $count_i$ records $\widehat{Cov}(S_i)$ (Ln 1). Then we scan each entry, if the value corresponds to the entry is not provided by $\mathbb{S}$ (flag.bit is false), for each source $S_i$ containing in this entry, we update $\widehat{Cov}(S_i)$ by accumulating $count_i$ with the truth probability of value stored in this entry (Ln 2-5). After that, we have obtained $\widehat{Cov}(S_i)$ for each $S_i \in \Omega \setminus \mathbb{S}$, and we select one (called $S^*$) with the highest selection ratio (Ln 6). Finally, for each entry containing $S^*$, we set its flag bit to be true, denoting that the value corresponding to this entry will not be accessed in later rounds (Ln 7-9).

Continuing with the motivating example, we show the process of selecting $S^*$ in the first round and second round.

*Example 5:* In the first round, we start with an empty set $\mathbb{S}$. Table III reports the irreplaceable contribution of data sources in the first round. For convenience, we set the cost of each data source be 1. For each source $S$, we compute $\widehat{Cov}(S)$ respectively. We scan each entry to find all the presences of $S_1$, $AT\&T.DA(.64)$, $Google.CA(.74)$, $MS.WA(.99)$, $IBM.WA(.34)$ and $Apple.CA(.93)$. Next, we accumulate truth probability of each value to get $\widehat{Cov}(S_1) = .64 + .74 + .99 + .34 + .93 = 3.64$. Among all the sources, $S_1$ owns the highest selection ratio (3.64). We select it and set all the entries flag to be true with the presence of $S_1$. (Shown in Table IV)

Then we consider the second round. Table V lists the irreplaceable contribution for the remaining data sources. Consider $S_2$, all the presences of $S_2$ in the remaining entries are, $Google.LA(.26)$ and $IBM.NY(.58)$ (AT&T.DA, MS.WA, Apple.CA are covered by $S_1$). Accumulating them we obtain $\widehat{Cov}(S_2)(.84)$. It is exactly the same way for computing other $\widehat{Cov}(S)$. Finally in the second round, $S_2$ with the highest selection ratio is selected.

TABLE III. IRREPLACEABLE CONTRIBUTION IN FIRST ROUND

| source | irreplaceable contribution |
|---|---|
| $S_1$ | .64+.74+.99+.34+.93=3.64 |
| $S_2$ | .64+.26+.99+.58+.93=3.40 |
| $S_3$ | .64+.58+.93=2.15 |
| $S_4$ | .23+.26+.99+.08+.93=2.49 |
| $S_5$ | .74+.01+.58+.02=1.35 |
| $S_6$ | .64+.74+.01+.58=1.97 |
| $S_7$ | .13+.74+.99+.58+.05=2.49 |

Next we analyze the time complexity of *Selectmax*.

We denote the queried data item set by $\overline{D}$. The set of values describing $\overline{D}$ is $\overline{V}$ with its cardinality as $|\overline{V}|$, which is also the number of queried entries in the index. From Table II, $|\overline{V}|$ is 14 in the motivating example. Recall that $|\Omega|$ is the number of sources which are present in the index.

In each round, we need to calculate $\widehat{Cov}(S)$ for each $S \in \Omega \setminus \mathbb{S}$. Thus we scan at most $|\overline{V}||\Omega|$ records in index because the number of entries is $|\overline{V}|$ and each entry contains at most $|\Omega|$ sources. Then we could draw the complexity as $O(|\overline{V}||\Omega|)$.

TABLE IV. SECOND ROUND OF INDEX

| entry | flag | corresponding sources | probability |
|---|---|---|---|
| **AT&T.DA** | T | $S_1, S_2, S_3, S_6$ | .64 |
| AT&T.TE | F | $S_4$ | .23 |
| AT&T.NY | F | $S_7$ | .13 |
| **Google.CA** | T | $S_1, S_5, S_6, S7$ | .74 |
| Google.LA | F | $S_2, S_4$ | .26 |
| **MS.WA** | T | $S_1, S_2, S_4, S_7$ | .99 |
| MS.BJ | F | $S_5, S_6$ | .01 |
| MS.TX | F | $S_3$ | 0 |
| IBM.NY | F | $S_2, S_3, S_5, S_6, S_7$ | .58 |
| **IBM.WA** | T | $S_1$ | .34 |
| IBM.BS | F | $S_4$ | .08 |
| **Apple.CA** | T | $S_1, S_2, S_3, S_4$ | .93 |
| Apple.WA | F | $S_7$ | .05 |
| Apple.NY | F | $S_5$ | .02 |

TABLE V. IRREPLACEABLE CONTRIBUTION IN SECOND ROUND

| source | $S_2$ | $S_3$ | $S_4$ | $S_5$ | $S_6$ | $S_7$ |
|---|---|---|---|---|---|---|
| $\widehat{Cov}(S)$ | .84 | .00 | .58 | .03 | .59 | .76 |

*C. Greedy Selection*

In this subsection, we present two greedy-based algorithms for source selection. Then we show the analysis of complexity and prove the approximations of both algorithms.

As the initialization, we load the entries in the index related to the query and take the data sources in these entries as the input. Note that the sources in other entries could not provide results for the query and can be excluded safely.

The two greedy algorithms for two problem, MaxContribution and MinCost, share the same framework of greedy algorithm: in each iteration, it selects a source with the highest selection ratio.

*MinCost* algorithm is shown in Algorithm 2. As for the *MaxContribution*, the only difference is that we replace the while condition with ($\exists S \in \Omega \setminus \mathbb{S}$ such that $c(\mathbb{S}) + c_S \leq L$). Next we focus on *MinCost* algorithm. In each iteration, we select a source $S$ with the highest selection ratio from the unselected source set (Ln 3). Then we join $S$ into the selected sources set $\mathbb{S}$ and update the answer set accordingly (Ln 4). The termination condition is that no more sources can provide new values to the selected answer set $\overline{V}$ (Ln 2).

If the constraint condition *cost* is appended, algorithm stops iff. one of following conditions is satisfied. (1) the selected sources $\mathbb{S}$ have already provided all answers; or (2) the total cost exceeds budget.

Next, we show the time complexity of greedy-based algorithms in Proposition 1.

*Proposition 1:* Time complexity of the greedy-based algorithm is $O(|\overline{V}|^2|\Omega|)$.

*Proof:* From aforementioned analysis, the time complexity of one iteration is $O(|\overline{V}||\Omega|)$. (In Section III-B)

Assume that finally we select $r$ sources, and let $n_i$ be the number of values covered by source $S_i$ in the $i$-$th$ round. Clearly, $\sum_{i=1}^{r} n_i = |\overline{V}|$, since in each round we choose one source. Thus, we have $r \leq |\overline{V}|$, since $n_i \geq 1, i = 1, 2, ..., r$. Finally, the overall time complexity of the algorithm is $O(|\overline{V}|^2|\Omega|)$. ∎

**Algorithm 1** Selectmax
**Input:**
  Sources set $\Omega$, selected sources set $\mathbb{S}$, cost function $c$.
**Output:**
  $S^*$ // source with the highest selection ratio
1: $count_i \leftarrow 0;$ //$count_i$ records $\widehat{Cov}(S_i)$
2: **for** each entry $E$ **do**
3:   **if** $flag.bit = false$ **then**
4:     **for** each source $S_i$ containing in the entry **do**
5:       $count_i \leftarrow count_i + P(v);$
6: $S^* = argmax_{S_i \in \Omega \setminus \mathbb{S}} count_i / c_{S_i};$
7: **for** each entry **do**
8:   **if** $flag.bit = false$ & $S^*$ is contained in the entry **then**
9:     $flag.bit = true;$
10: **return** $S^*;$

**Algorithm 2** MinCost
**Input:**
  Sources set $\Omega$, cost function $c$.
**Output:**
  Selected sources set $\mathbb{S}$.
1: $\mathbb{S} = \emptyset, \overline{V} = \emptyset;$ //$\overline{V}$ is the set of values provided by $\mathbb{S}$, $V(S)$ is the set of values provided by $S$;
2: **while** ($\exists S \in \Omega \setminus \mathbb{S}$ such that $V(S) \nsubseteq \overline{V}$) **do**
3:   $S^* \leftarrow$ Selectmax$(\Omega, \mathbb{S}, c);$
4:   $\mathbb{S} \leftarrow \mathbb{S} \cup \{S^*\}, \overline{V} \leftarrow \overline{V} \cup V(S^*);$
5: **return** $\mathbb{S}$.

Note that these two algorithms, MaxContribution and MinCost, share the same time complexity, since the worst case of the former is the same as the latter when cost budget exceeds the sum cost of all data sources. In practice, greedy algorithm leveraging index can achieve high performance on massive datasets since $|\Omega|$ is usually large while $|\overline{V}|$ is small.

Next, we show the approximation of MinCost and MaxContribution and prove both of them.

*Theorem 2:* Let $\alpha$ denote the number of values in the largest source in the input of MinCost problem. MinCost algorithm achieves a $log\alpha$-factor approximation to the optimal solution.

*Proof:* We prove it by showing that there exists a L-reduction from MinCost problem to set cover problem with the universal set $U$ and the family of subsets $\mathfrak{S}=\{s_1, s_2, ..., s_n\}$.

Given an instance $I_{min}$ of MinCost with a source set $\Omega=\{S_1, S_2, ..., S_n\}$, each source $S_i$ with a cost $c_{S_i}$, and $V$ as the set of all values, we construct an instance $I'_{min}$ for set cover problem as follows. As the reduction, the universal set $U$ is $V$ and the family of subsets is $\mathfrak{S}$ such that each source $S_i$ corresponds to a subset of $s_i \subseteq U$, where $s_i$ contains the elements corresponding to the values in $S_i$. The weight of $s_i$ is the cost of $S_i$.

Given an arbitrary feasible solution of set cover problem, denoted by $R=\{s_i : s_i \in \mathfrak{S}\}$, we construct a solution of MinCost denoted by $D$ with $|R|=|D|$ as follows. First, for each $s_i \in R$, the values covered by $S_i \in D$ correspond to the elements covered by $s_i$. Second, the cost of $S_i$ equals the weight of $s_i$. We denote by $OPT_s$ and $OPT_m$ the optimal solution of set cover problem and MinCost problem, respectively. Thus, we have $OPT_s(I'_{min}) \leq OPT_m(I_{min})$ and $|OPT_m - |D|| \leq |OPT_s(I'_{min}) - |R||$.

The reduction is accomplished in polynomial time and the algorithm solving set cover problem yields an approximation ratio of $log\epsilon$, where $\epsilon$ is the number of elements in the largest subset. Thus, MinCost mimics the greedy algorithm to obtain an approximation of $log\alpha$, where $\alpha$ is the number of values in the largest source. ∎

*Theorem 3:* The approximation ratio bound of MaxContribution algorithm is $1 - \frac{1}{e}$.

*Proof 2:* We first prove that MaxContribution is equivalent to the maximum coverage problem (MCP) [10].

Given an instance of $I_{max}$ of MaxContribution with a source set $\Omega = \{S_1, S_2, ..., S_n\}$, each source $S_i$ with a cost $c_{S_i}$, and $V = \{v_1, v_2, ..., v_m\}$ as the set of all values, each value $v_i$ with a truth probability $p_i$. We construct an instance $I'_{max}$ as follows: let $\mathcal{S}$ be $\Omega$ and elements domain $X$ be value set $V$. The cost of $S_i \in \mathcal{S}$ corresponds to that of $S_i \in \Omega$ and the weight of element $x_i$ corresponds to truth probability of $v_i$. Given a solution of MCP, denoted by $R = \{s_i : s_i \in \mathcal{S}\}$. Then we can construct a solution of MaxContribution denoted by $D$ with $|R| = |D|$ as follows. For each $s_i \in R$, the values covered by $S_i \in D$ correspond to the elements covered by $s_i$. The cost of $S_i$ equals the cost of $s_i$. Thus MaxContribution can be reduced to MCP. Conversely, MCP can be reduced to MaxContribution, which can be proved in the same way.

The $(1 - 1/e)$-approximation is based on reducing from MCP to the k-Coverage problem. [11].

## IV. OPTIMIZATION

In this section, we improve the efficiency of aforementioned algorithms. We start with discussion about the opportunities for improvement and then propose two pruning strategies.

First, for a data source, if the truth probabilities of some values are low, but those of the remaining values are still high, we should not abandon the data source but just exclude the low-truth-probability values during source selection. We capture this intuition by *Value-level* pruning.

Second, in one iteration the index algorithm takes $O(|\overline{V}||\Omega|)$ steps to select a data source with the highest selection ratio. We can reduce the time cost of one iteration by reducing the computational complexity of examining one source. For a data source, if the upper bound of its selection ratio is less than the best ratio probed so far, it can be pruned. We capture this intuition by *Source-level* pruning.

Then, we will introduce *Value-level* and *Source-level* pruning strategies in Section IV-A and Section IV-B, respectively.

### A. Value level pruning

In this subsection, we seek to prune those values with low truth probability. Specifically, we set a probability threshold $p$. For the values corresponding to the same queried data item, if the sum of truth probabilities of some values is smaller than $p$, they are treated as low-truth-probability values and could be pruned.

Note that this approach is equivalent to that of simply pruning values with the truth probability smaller than a threshold $p'$. Since if we seek to prune values with truth probability smaller than $p'$, we can always obtain the same result by modifying $p$. We choose such approach for the convenience of the proof of accuracy.

According to above discussions, this approach could be implemented by sorting the values corresponding to a data item in ascending order of truth probability and keeping pruning values until the sum of truth probability of the pruned values exceeds $p$.

Next we show that with such pruning strategy, we can prune low-truth-probability values without sacrificing accuracy much.

Specifically, for a data item $D$, we denote by $D(V)$ and $\widehat{D}(V)$ the set of values and the set of pruned values corresponding to $D$, respectively. First, for each value $v \in D(V)$, we sort it in ascending order according to its truth probability. Second, we set the threshold $p$ and prune such values that $\sum_{v \in \widehat{D}(V)} P(v) \leq p$. Next, let $\varrho$ be the probability that the remaining values still provide true one even if $\widehat{D}(V)$ are omitted. Note that in practice we can assign the value of $\varrho$ and next we will show the upper bound of $p$ corresponding to a fixed value $\varrho$.

Recall that in Section II-A we assume that among all distinct values describing a data item, only one is true and the rest are false. Before describing *Value-level* strategy, we first formulate some key notions. We denote by $P(\mathbb{V} \propto v)$ the probability that true value is in the remaining value set $\mathbb{V} = D(V) \setminus \widehat{D}(V)$. That is, the probability that one of values in $\mathbb{V}$ is true, which is equivalent to the probability that each value in $\widehat{D}(V)$ is false. Thus we can devise this notion as follows,

$$P(\mathbb{V} \propto v) = \prod\nolimits_{v \in \widehat{D}(V)} (1 - P(v)) \quad (8)$$

Intuitively, it is desirable that the pruned values would not sacrifice the accuracy of our final results much. That is, with values in $\widehat{D}(V)$ having been pruned, $P(\mathbb{V} \propto v) \geq \varrho$ still holds conditioned on a high value $\varrho$. Next, we will devise the boundary of $p$ after $\varrho$ is fixed. Applying the inequality of geometric mean and arithmetic mean,

$$\left(\prod\nolimits_{i=1}^{n} x_i\right)^{\frac{1}{n}} \leq \frac{\sum_{i=1}^{n} x_i}{n} \quad (9)$$

Applying above inequality leads us to devise the boundary of $p$ as follows,

$$\varrho \leq \prod\nolimits_{v \in \widehat{D}(V)} (1 - P(v)) \leq \left(\frac{\sum_{v \in \widehat{D}(V)} (1 - P(v))}{n}\right)^n$$
$$= \left(1 - \frac{\sum_{v \in \widehat{D}(V)} P(v)}{n}\right)^n = \left(1 - \frac{p}{n}\right)^{\frac{n}{p} \cdot p}$$
$$\leq e^{-p} (n \to \infty) \quad (10)$$

In above derivation process, we abuse the notion and denote $n = |\widehat{D}(V)|$. According to Equation 10, we draw that

*Theorem 4:* If $\varrho$ is fixed, we can bound $p$ as follows,

$$p \leq ln(\frac{1}{\varrho}) \qquad (11)$$

Theorem 4 can be easily devised according to Equation 10. Note that we transform the condition $\sum_{v \in \widehat{D}(V)} P(v) \leq p$ to $\sum_{v \in \widehat{D}(V)} P(v) = p$ in derivation process, and this will not exert an influence on the computation for upper bound of $p$. In practice, we compute $p$ by Equation 11 and prune values until its truth probability accumulation exceeds $p$. For example, if we set $\varrho=0.8$, $p = ln(\frac{1}{0.8}) = 0.22$. Thus we prune values until the sum of their truth probability exceeds 0.22. In this case, we still believe that probability that the true value lays in the remaining values is no less than 0.8.

In the process of source selection, we fix $\varrho$ as a constant all the time or adjust it in different rounds. Note that according to Equation 11 the upper bound of $p$ monotonically decreases with $\varrho$. That is, the higher $\varrho$ is, the lower $p$ turns, and the less values are pruned, which coincides with our intuition.

*Example 6:* Continue with motivating example in Table II, three values are provided for data item MS, WA(.99), BJ(.01) and TX(.0) respectively. If we fix $\varrho$ as 0.9, $p = ln(\frac{1}{0.9}) = 0.11$. Thus, BJ and TX are pruned. Next we consider another case when data item is Apple and it is described by CA(.93), WA(.05) and NY(.02). We set $\varrho$ as 0.93 and thus $p = ln(\frac{1}{0.93}) = 0.073$. Thus we have a high confidence 0.93 even if WA and NY have been pruned.

Let $|A|$ be the number of unpruned values we present the complexity of value level pruning in Proposition 2.

*Proposition 2:* Computational complexity of the algorithm with value level pruning is $O(|A|^2|\Omega|)$.

*Proof 3:* Recall that $|A|$ is the number of values which are not pruned. With value level pruning strategy, the number of entries decreases to $|A|$.

In each round, for each unselected source, we need to calculate its irreplaceable contribution and choose the source with the highest selection ratio. Note that in the inverted index, there are at most $|A||\Omega|$ records, since the number of entries is $|A|$ and each entry contains at most $|\Omega|$ sources. After scanning all sources for the queried entries, we obtain $\widehat{C}(S)$ for each unselected source $S$ and select the one with the highest ratio. Thus, the time complexity of one round is $O(|A||\Omega|)$.

Assume that finally we select $r$ sources, and let $n_i$ be the number of values covered by source $S_i$ in $i$-th round. Clearly, $\sum_{i=1}^{r} n_i = |A|$, since in each round we choose one source. Thus, we have $r \leq |A|$, since $n_i \geq 1, i = 1, 2, ..., r$. Finally, the overall time complexity of the algorithm is $O(|A|^2|\Omega|)$.

### B. Source level pruning

In this part, we further improve the efficiency of index algorithm by reducing the time cost of examining data sources in iterations.

In one iteration, we select a data source with the highest selection ratio. That is,

$$S^* = argmax_{S \in \Omega \setminus \mathbb{S}} \widehat{Cov}(S)/c_S \qquad (12)$$

The intuition of *Source-level* is that for each data source $S$, we tend to compute an upper bound for $\widehat{Cov}(S)/c_S$ in constant time. If its upper bound is less than the best selection ratio probed so far, we can prune $S$.

Now the questions are:

*1) How to compute the upper bound?*

*2) How can we maintain the upper bound in constant time?*

In the remainder of this subsection, we begin with describing the computational method of upper bound and then illustrate how to maintain the upper bound in constant time by algorithm pseudo code.

For a data source $S$, we compute the upper bound of selection ratio $\widehat{Cov}(S)/c_S$. Note that the cost of $S$, $c_S$, is known as the input of algorithm. Thus we only need to calculate the upper bound of $\widehat{Cov}(S)$. Recall that given the set of selected source set $\mathbb{S}$, $\widehat{Cov}(S)$ denotes the irreplaceable contribution of $S$ (See Equation 7), which is the sum of truth probability of those values *only* provided by $S$. In other words,

$$\widehat{Cov}(S) = Cov(\mathbb{S} \cup \{S\}) - Cov(\mathbb{S}) \qquad (13)$$

where $Cov(\mathbb{S})$ is the contribution of source set $\mathbb{S}$, denoting the sum of truth probability of all distinct values provided by $\mathbb{S}$ (See Equation 2). Thus, we turn to compute the upper bound of $Cov(\mathbb{S} \cup \{S\}) - Cov(\mathbb{S})$. Before discussing the bound, we first prove the submodularity of set function $Cov(.)$ to give the theoretical guarantee [12].

*Definition 7:* (Submodularity) Given a finite set $\Omega$, a set function $f$: $2^{|\Omega|} \rightarrow \mathcal{R}$ is submodular iff. for any two sets $\mathcal{S} \subseteq \mathcal{T} \subseteq \Omega$ and $S \in \Omega \setminus \mathcal{T}$, it holds that $f(\mathcal{S} \cup \{S\}) - f(\mathcal{S}) \geq f(\mathcal{T} \cup \{S\}) - f(\mathcal{T})$.

*Lemma 1:* Given a finite set $\Omega$, $Cov(.)$ is a submodular set function.

*Proof:* For any two sets of data sources $\mathcal{S} \subseteq \mathcal{T} \subseteq \Omega$ and $S \in \Omega \setminus \mathcal{T}$, it derives that,

$$Cov(\mathcal{S} \cup \{S\}) - Cov(\mathcal{S}) = \sum_{v \in (Q(S) - Q(S \cap S))} P(v)$$

Since $\mathcal{S} \subseteq \mathcal{T}$, $(\mathcal{S} \cap S) \subseteq (\mathcal{T} \cap S)$. Thus, we have $Q(\mathcal{S} \cap S) \subseteq Q(\mathcal{T} \cap S)$. Then we can deduce that $Q(S) - Q(\mathcal{S} \cap S) \supseteq Q(S) - Q(\mathcal{T} \cap S)$. Finally we have $\sum_{v \in (Q(S) - Q(S \cap S))} P(v) \geq \sum_{v \in (Q(S) - Q(S \cap T))} P(v)$, that is,

$$Cov(\mathcal{S} \cup \{S\}) - Cov(\mathcal{S}) \geq Cov(\mathcal{T} \cup \{S\}) - Cov(\mathcal{T})$$

Thus we conclude that given a set of data sources $\Omega$, $Cov(.)$ is a submodular set function. ∎

*Theorem 5:* Given a set of data sources $\Omega$ and any two source set, $\mathbb{T} \subseteq \mathbb{S} \subseteq \Omega$. For each data source $S \in \Omega \setminus \mathbb{S}$, $Cov(\mathbb{S} \cup \{S\}) - Cov(\mathbb{S}) \leq min(Cov(\Omega) - Cov(\mathbb{S}), Cov(\mathbb{T} \cup \{S\}) - Cov(\mathbb{T}))$.

*Proof:* First, clearly for any source set $\mathbb{S} \subseteq \Omega$ and any source $S \in \Omega \setminus \mathbb{S}$, $Cov(\mathbb{S} \cup S) \leq Cov(\Omega)$. Thus we have $Cov(\mathbb{S} \cup \{S\}) - Cov(\mathbb{S}) \leq Cov(\Omega) - Cov(\mathbb{S})$.

Second, according to Lemma 1, we can deduce that $Cov(\mathbb{S} \cup \{S\}) - Cov(\mathbb{S}) \leq Cov(\mathbb{T} \cup \{S\}) - Cov(\mathbb{T})$.

Finally, we conclude that $Cov(\mathbb{S} \cup \{S\}) - Cov(\mathbb{S}) \leq min(Cov(\Omega) - Cov(\mathbb{S}), Cov(\mathbb{T} \cup \{S\}) - Cov(\mathbb{T}))$. ∎

From Theorem 5, we draw that given the selected source set $\mathbb{S}$ and any source set $\mathbb{T} \subseteq \mathbb{S}$, for any unselected source $S$, the upper bound of $\widehat{Cov}(S)/c_S$, denoted by $Up(S, \mathbb{S})$, is,

$$Up(S, \mathbb{S}) = min(Cov(\Omega) - Cov(\mathbb{S}), Cov(\mathbb{T} \cup \{S\}) - Cov(\mathbb{T}))/c_S$$

Next we illustrate the *Source-level* pruning algorithm and discuss how to obtain the upper bound in constant time.

TABLE VI. VARIABLES IN ONE ITERATION

| Variable | Meaning |
|---|---|
| $r$ contribution | $\widehat{Cov}(S)/c_S$, the ratio of irreplaceable and cost for source $S$. |
| $Con(S, \mathbb{S})$ contribution | $Cov(\mathbb{S} \cup \{S\}) - Cov(\mathbb{S})$, irreplaceable of $S$ with respect to $\mathbb{S}$. |

**Algorithm 3** Source-level Prune in one Iteration

**Input:**
  Source set $\Omega$, selected source set $\mathbb{S}$, cost function $c$.
**Output:**
  $S^*$. // Source with the highest $\widehat{Cov}(S)/c_S$.
1: $r \leftarrow 0$, $r^* \leftarrow 0$;
2: **for** each $S \in \Omega \setminus \mathbb{S}$ **do**
3:   **if** $Up(S, \mathbb{S}) < r^*$ **then**
4:     **continue**; // prune
5:   $\widehat{Cov}(S) \leftarrow Cov(\mathbb{S} \cup \{S\}) - Cov(\mathbb{S})$;// Update using Equation 14
6:   $Con(S, \mathbb{S}) \leftarrow \widehat{Cov}(S)$;
7:   $r \leftarrow \widehat{Cov}(S)/c_S$;
8:   **if** $r > r^*$ **then**
9:     $r^* \leftarrow r$, $S^* \leftarrow S$;
10: **return** $S^*$;

For each unselected data source $S$, if its upper bound is less than the highest ratio $r^*$ probed so far, it can be pruned (Ln 3-5). Then we update the irreplaceable contribution of $S$ by Equation 14, and we store the difference between $Cov(\mathbb{S} \cup \{S\})$ and $Cov(\mathbb{S})$ in variable $Con(S, \mathbb{S})$ in each iteration (Ln 6-7). If $S$ has not been pruned, $r$ turns to be a tighter bound of $S$ since $r$ is updated to a smaller number than that in any previous iterations according to Lemma 1. Next, if the ratio $r$ of $S$ is higher than the best ratio so far, we updated $r^*$ and $S^*$ with $r$ and $S$, respectively (Ln 8-11). In one iteration, given the selected source set $\mathbb{S}$, when we compute the upper bound for a data source $S$, the variable $Con(S, \mathbb{T})(\mathbb{T} \subseteq \mathbb{S})$ has been stored in the previous iterations and we can obtain it in constant time. Based on above analysis, we can compute the upper bound of a data source in constant time.

Note that source level pruning strategy will not sacrifice the accuracy of results since we only prune those sources whose upper bound is less than the best ratio probed so far.

Let $|\mathcal{U}|$ be the average number of sources in each iteration that are not pruned. We present the overall complexity analysis in Proposition 3.

*Proposition 3:* Computational complexity of algorithm with source level and value level pruning strategies is $O(|A|^2|\mathcal{U}| + |A||\Omega|)$.

*Proof 4:* Recall that $|\mathcal{U}|$ is the average number of sources in each iteration that are not pruned.

In each iterations, we need to store the difference between $C(\mathbb{S} \cup \{S\})$ and $C(\mathbb{S})$ in variable $Con(S, \mathbb{S})$ for each unselected data source. From the analysis of Proof 3 we can deduce that there are at most $|A|$ iterations. And in each iteration we store $Con(S, \mathbb{S})$ value for at most $|\Omega|$ data sources. Thus, time complexity of maintaining the upper bound is $O(|A||\Omega|)$.

As for source selection process, in each round, a source can be pruned in constant time if it meets the pruning condition. And for the remaining sources which are not pruned, whose cardinality is $|\mathcal{U}|$, we apply the same analysis with Proof 3. Thus computational complexity for source selection process is $O(|A|^2|\mathcal{U}|)$.

Overall, time complexity of index algorithm with source level and value level pruning strategies is $O(|A|^2|\mathcal{U}| + |A||\Omega|)$.

## V. EXPERIMENT

This section presents experimental results on both synthetic and real-world datasets validating the efficiency and effectiveness of the proposed algorithms as well as pruning strategies in this paper.

### A. Experiment setup

Data: To test the performance completely, we use two datasets.

The *Book* dataset contains 1263 books and 894 data sources (bookstores). Information on Computer Science books was collected from online bookstore aggregator AbeBooks.com in 2007. We use the golden standard that provides precise author lists manually obtained from book covers on 100 randomly selected books. We match our selected results with the golden standard to calculate the recall of our methods.

To test our methods with various parameters settings when source volume scales to million, we execute algorithms and pruning strategies on synthetic dataset. Next we describe how to generate synthetic data in the following aspects:

(1) Data Source: For scalability purpose, we generate the number of sources ranging from one million to ten million. To simulate real data sources, we assume that the number of data items contained in one source follows normal distribution (denoted by $\mathcal{N}_S$). To generate sufficient large data sources and ensure the variance of the sizes of data sources, we set mean value $\mu$=100,000 and $\sigma=\mu/3$.

(2) Data Errors: To simulate various source accuracy in real world, we generate errors into each data source and the number of errors for each source is generated following normal distribution whose mean value $\mu'$ is proportional to that of distribution $\mathcal{N}_S$ and $\sigma' = \mu'/3$.

(3) Golden Standard: On synthetic dataset, we decide true value for any data item manually and randomly select a set of data items as the golden standard.

Implementation: Recall that our proposed algorithms, Max-Contribution and MinCost share the same framework and the only difference is the stop condition. MaxContribution stops iff one of following conditions is satisfied: (1) selected sources have already provided all answers; or (2) the total cost exceeds budget. As for MinCost, it stops as long as condition (1) is satisfied. Thus, except *cost budget* factor, the other experimental results for these two algorithms are exactly the same and we use the term *basic greedy algorithm* to refer to MaxContribution and MinCost in the remainder of this section. And we use the algorithm $GRASP$ proposed by [1] as state-of-art for comparison.

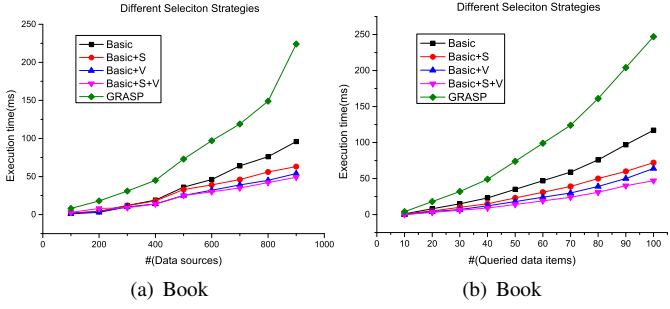

(a) Book  (b) Book

Fig. 1. Efficiency Test on Book

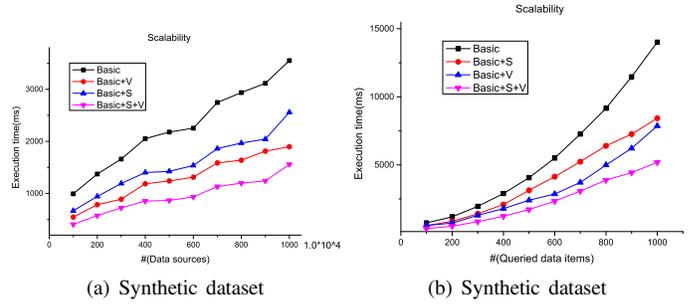

(a) Synthetic dataset  (b) Synthetic dataset

Fig. 2. Efficiency Test on Synthetic Dataset

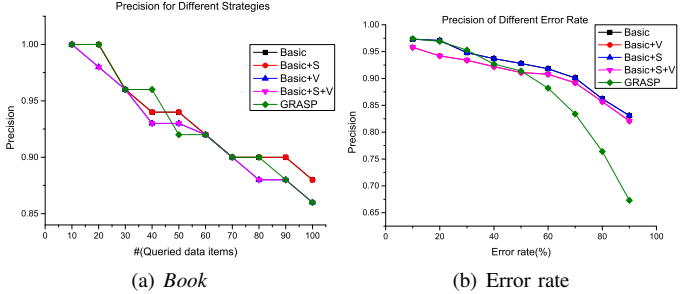

(a) *Book*  (b) Error rate

Fig. 3. Precision Test for Various Strategies

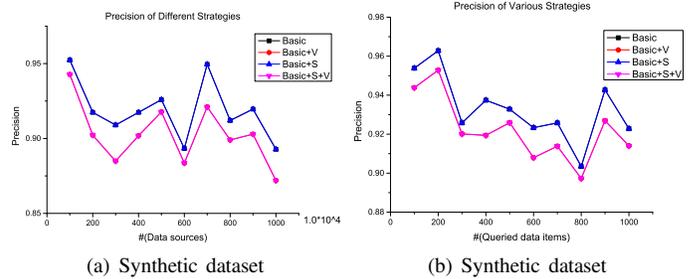

(a) Synthetic dataset  (b) Synthetic dataset

Fig. 4. Precision Test for Various Strategies

In this section, we implement five algorithms: the basic greedy algorithm (*Basic*), algorithm with source level pruning (*Basic+S*), algorithm with value level pruning (*Basic+V*), algorithm with both source and value level prunings (*Basic+S+V*) and *GRASP*.

We implement all algorithms in C++ on a Windows machine with Intel Core i3 (2.1GHz, 4GB memory).

Measures: As for selection results, we use recall and precision to measure the quality of our selected sources. *Precision* is defined as the ratio of the number of true values and all values provided by selected source set. And *recall* is evaluated as the proportion of golden standard covered by answers from the selected sources. For efficiency test, we count execution time for various methods.

### B. Efficiency and Scalibility

First, to test how the number of sources affects efficiency, we compare five strategies on *Book* dataset and the results are shown in Fig 1(a). We observe that (1) as we query more sources, execution time of five strategies increases gradually. Among them, *GRASP* has the lowest performance. On average, it takes more time than *Baisc* by 1.9 times, than *Basic+S* by 3 times, than *Basic+V* by 3.5 times and than *Basic+S+V* by 3.8 times, respectively. (2) As for the four proposed strategies, we sort them in ascending order of runtime: *Basic+S+V*, *Basic+V*, *Basic+S* and *Basic*. The slow growth shows great efficiency.

Then, we study the effect of data item size. We randomly choose data items from golden standard and range the number from 10 to 100 step by 10. As shown in Fig 1(b), greedy-based strategies leveraging index outperform *GRASP*. Among four proposed strategies, with the presence of source-level pruning, running time climbs more slowly than the others and the difference of time cost between *Basic+S* (finished in 72ms) and *Basic+V* (finished in 64ms) is getting smaller.

Second, we consider scalability and conduct two experiments on synthetic datasets. Fig 2(a) plots the execution time as we vary the number of data sources from one million to ten million. Since the run time of *GRASP* on synthetic dataset is over three days, we omit it in the scalability comparison. As we query sources, execution time for four strategies increases nearly linearly. Compared with *Basic*, on average *Basic+S* beats it by 30%, *Basic+V* reduces time by 46% and *Basic+S+V* by 56%. Among them, *Basic+S+V* achieves the best performance. It finishes in 1557ms when the source number reaches ten million, showing that pruning techniques improve basic algorithm significantly. Next, we examine how the number of queried data items affects efficiency on synthetic dataset in Fig 2(b). We start with 100 data items and add items until reaching 1000. Our observations are consistent with those in Fig 1(b): (1) For strategies leveraging source-level pruning, *Basic+S* and *Basic+S+V*, runtime increases linearly w.r.t the size of queried data items while the other two strategies climbs quadratically; (2) Even in dataset with up to ten million sources, *Basic+S+V* finishes within 5 seconds and *Basic* takes less than 15 seconds, showing the great scalability of our methods.

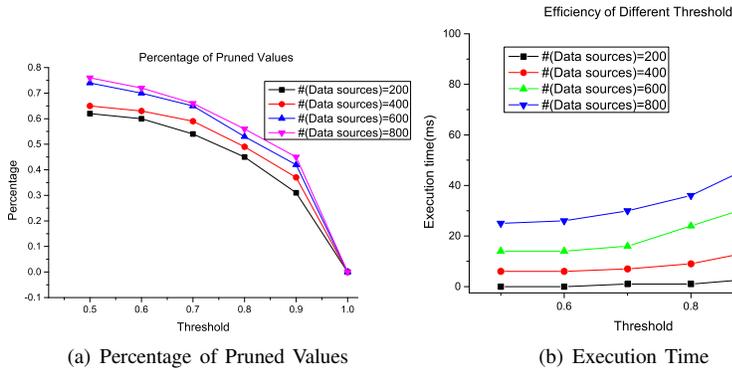

(a) Percentage of Pruned Values

(b) Execution Time

Fig. 5. Prune Ratio for Value-level

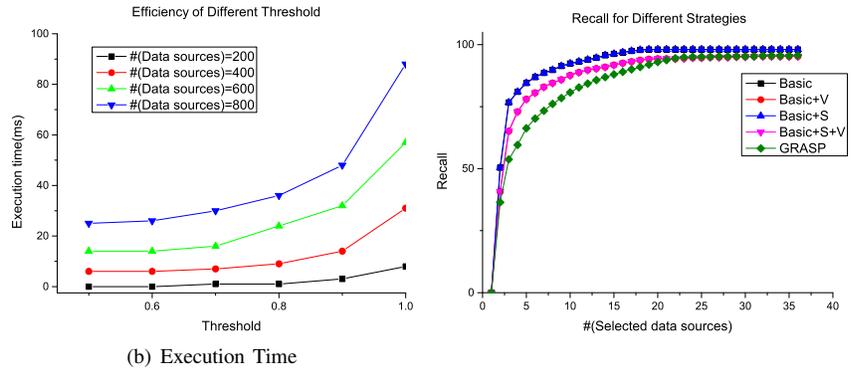

Fig. 6. Recall Test on Book

### C. The Quality of Selected Sources

**Precision:** We conduct four experiments on precision of results conditioned on various parameters.

First, we study the precision of various strategies when varying the number of queried data items from 10 to 100 on *Book* dataset. From Fig 3(a), we observe that: (1) As we query more data items, precision of all strategies drops in general. For *Baisc*, precision remains 1 till 20 data items and then decreases to 0.88 till 100 data items. (2) The curves of *Basic* and *Basic+S* coincide, so do *Basic+V* and *Basic+S+V*, showing that source-level pruning will not sacrifice the precision of result. (3) On average *Basic* beats *GRASP* slightly by one percent while *Basic+S+V* lowers precision by two percent, showing that value-level will not sacrifice precision much, and our proposed methods have the similar high precision with *GRASP*.

Second, to test the effectiveness when scaling the data volume to million level, we show the results on synthetic dataset in Fig 4. Fig 4(a) and Fig 4(b) plot precision conditioned on the various number of data sources and queried data items, respectively. For Fig 4(a), we range the number of sources from one million to ten million. Each time we generate data source set $\Omega$ randomly. Note that source set with different sizes is generated independently. And we observe that: (1) No obvious tendency for precision changes monotonically with the increase of source number and precision ranges from .89 to .95, showing that our methods still achieve high precision when source number scales up; (2) the precision of *Basic* and *Basic+S* are exactly same, so do *Basic+V* and *Basic+S+V*, which is consistent with the observations on Fig 3(a). For Fig 4(b), we generate ten sets of queried data item set, whose size ranges from 100 to 1000. Note that each item set is generated independently. When we query more data items, precision fluctuates in the range of .89 to .96. Both experiments on synthetic dataset show that our methods can still have a high precision on dataset with up to millions of sources.

Finally, we examine the effect of data error by generating different fractions of errors into data (called error rate). As shown in Fig 3(b), the precision of *Basic+S* and *Basic* are exactly the same, so are *Basic+V* and *Basic+S+V*, showing that source-level pruning will not sacrifice precision. As we increase error rate from $10\%$ to $90\%$, precision gradually drops until reaching around .82 and strategies leveraging value-level loss precision by 1.8 percent on average. Note that even if error rate reaches $90\%$, precision can still be over .8, showing the robustness of our strategies. Regarding the precision of *GRASP*, it declines sharply as the increase of error rate, from above $90\%$ to below $60\%$. Thus we can draw that our proposed methods could handle data with low error rate.

**Recall:** We compare five strategies: *Basic*, *Basic+S*, *Basic+V*, *Basic+S+V* and *GRASP* on *Book* dataset to show recall of source selection results. As shown in Fig 6, the x-axis represents the number of selected data sources and we have the following observations. (1) Recall quickly increases and then flattens out for all strategies. (2) When we select 24 sources, *Baisc* and *Basic+S* converge. And the saturation point for *Basic+V* and *Basic+S+V* is 28, for *GRASP* is 34, showing that our proposed algorithms can stop earlier. (3) When algorithms converge, compared with *GRASP* (its recall is 0.95), *Baisc* and *Basic+S* beats it by 2.1 percent on average, while *Basic+V* and *Basic+S+V* are not performed as well as it (less than 1 percent on average).

### D. The Impact of Value-level Threshold

In this section, we conduct four experiments to examine how threshold $\varrho$ of *Value-level* affects the final results. Recall that in Section IV-A, for a data item $D$, $D(V)$ and $\widehat{D}(V)$ denote the set of values and the set of pruned values corresponding to $D$, respectively. And $\varrho$ is set by users (See Equation 11), denoting the probability that the remaining values still provide true one even if $\widehat{D}(V)$ are omitted.

First, to understand how the percentage of pruned values varies conditioned on different $\varrho$ and source number, we conduct experiments on *Book*. Note that when $\varrho$ and the number of sources are fixed, the percentage of pruned values varies for different data items and we report the average percentage in Fig 5(a). The x-axis represents $\varrho$, ranging from .5 to 1 and the y-axis represents the percentage of pruned values. We observe that: (1) When the number of data sources is fixed, the higher threshold is, the less percentage turns. When source number turns 200, we can prune nearly $75\%$ values with threshold being .5. As we increase $\varrho$ to .9, $45\%$ values can still be pruned. (2) As we add the number of data sources from 200 to 800, the percentage of pruned values climbs: from .62 to .76 when $\varrho$ is .5 and from .31 to .45 when $\varrho$ is .9. This is not surprising because the larger source number is, the more values provided for one data item are, and thus the higher percentage of low-truth-probability values will be pruned due to the assumption that only one value is true (See Section II). (3) As the number of data sources increases to 800, the percentage increases much more slowly and the difference between two curves (#(Data sources) is 600 and 800) is small. The slow growth is because given a fixed data domain, the values describing a specific data item is limited to a certain number. For *Book* dataset, we count the average number of values describing one data item in golden standard, which is 3.41. In other words, for a data item, there is one true value and 2.41 false values on average. Thus, the percentage of false values is around 70%. The above observations and discussions show that value-level pruning strategy can prune a large fraction of values even if $\varrho$ is set as a high value.

TABLE VII. PRECISION AND RECALL WITH DIFFERENT THRESHOLD

| Threshold | 0.5 | 0.6 | 0.7 | 0.8 | 0.9 |
|---|---|---|---|---|---|
| Precision | 0.87 | 0.87 | 0.87 | 0.87 | 0.87 |
| Recall | 0.93 | 0.94 | 0.96 | 0.96 | 0.96 |

Second, we study how threshold affects efficiency. As we increase threshold, execution time climbs gradually. When source number turns 800, compared with $\varrho = 1$ (without value-level pruning), $\varrho = .9$ saves time by 45%, $\varrho = .8$ by 59%, $\varrho = .7$ by 66%, $\varrho = .6$ by 70% and $\varrho = .5$ by 71%, respectively. We can draw that value-level pruning can speed up the basic algorithm significantly.

Finally, we test precision and recall of final results conditioned on various threshold value $\varrho$. We conduct experiments on *Book* and Table VII lists the results. In this test, we set $\varrho$ as .8 and fix the size of golden standard as 100. We query 893 sources on *Book* and observe that: (1) Precision remains unchanged, showing that threshold above .5 will not cause much effect on precision; (2) As we increase $\varrho$, recall declines slightly from .96 ($\varrho = .9$) to .93 ($\varrho = .5$), but not too much. The above observations show that our value-level pruning will not sacrifice the quality of source selection much.

**Summary:**

(1) our proposed methods can return a large portion of correct values; (2) compared with the state-of-art, our approaches can scale to datasets with up to millions of data sources while gain similar quality of selected results; (3) two pruning strategies can significantly improve the efficiency without sacrificing the quality of selected data sources much.

## VI. RELATED WORK

Studies about source selection are critical for data integration. Sources providing more relevant and true values for data items describing entities in real world are desirable to be selected in data fusion process. The work about offline or online source selection is not too much and recently [1], [2] studied source selection for various applications. [1] selected a set of sources efficiently to maximize the profit taking the monotonicity property into consideration and proposed a heuristic randomized approach. However, it could hardly scale to massive data sources (million level) due to the high time cost. [2] proposed a data integration technique with dependent sources under the assumption that all values provided by sources are true. Even though the proposed algorithm can estimate the coverage and select sources effectively, it does not consider data quality and owns high time complexity, which prevents the algorithm from scaling to massive data sources.

The evaluation of trustworthiness of sources is critical to the quality of selection results. [13] and [14] assign a global trust rating to each data source on a P2P network. Authority-hub analysis [15] and PageRank [16] decide trustworthiness based on link analysis [17]. As mentioned in Section II, our work is based on [3] to obtain the true probability of of values in data sources due to the following two reasons: [3] considers both source accuracy and copying relationship between sources, thus it can return a true probability with a high precision. Second, the computation process could be handled off-line efficiently.

Our work considers both efficiency and effectiveness. We propose a scalable algorithm with theoretical guarantee and two efficient pruning strategies which will not sacrifice the quality of selected data sources much.

## VII. CONCLUSIONS

This paper studies source selection problem taking efficiency and effectiveness into consideration. We first propose a probabilistic coverage model to evaluate the quality of data source and formulate source selection problem to two problems, which are both proven to be NP-hard. Then we describe the efficient computing method of quality metric leveraging index. Finally we develop greedy-based algorithms and show their approximations. To further improve efficiency, we propose two efficient pruning strategies. Both are with theoretical guarantee and will not sacrifice the quality of results much. Experimental results on both real world and synthetic datasets show our methods can select sources providing a large proportion of true values efficiently and can scale to datasets with up to millions of data sources.